\documentclass[conference]{IEEEtran}
\IEEEoverridecommandlockouts
\usepackage{cite}
\usepackage{amsmath,amssymb,amsfonts}
\usepackage{algorithmic}
\usepackage{graphicx}
\usepackage{textcomp}
\usepackage{xcolor}
\usepackage[hidelinks]{hyperref}
\usepackage{booktabs}
\usepackage{multirow}
\usepackage{caption}
\usepackage{siunitx}

\def\BibTeX{{\rm B\kern-.05em{\sc i\kern-.025em b}\kern-.08em
    T\kern-.1667em\lower.7ex\hbox{E}\kern-.125emX}}

\begin{document}

\title{Multi-Target Backdoor Attacks Against Speaker Recognition}

\author{
    \IEEEauthorblockN{\textit{Anonymous submission to ASRU 2025}}
}

\author{
\IEEEauthorblockN{
Alexandrine Fortier\IEEEauthorrefmark{1}\thanks{Corresponding author: alexf01@student.ubc.ca (work completed while at École de technologie supérieure)},
Sonal Joshi\IEEEauthorrefmark{2},
Thomas Thebaud\IEEEauthorrefmark{2},
Jesus Villalba Lopez\IEEEauthorrefmark{2},
Najim Dehak\IEEEauthorrefmark{2},
Patrick Cardinal\IEEEauthorrefmark{1}}
\IEEEauthorblockA{\IEEEauthorrefmark{1}Department of Software and IT Engineering, École de technologie supérieure, Québec, Canada}
\IEEEauthorblockA{\IEEEauthorrefmark{2}Department of Electrical and Computer Engineering, Johns Hopkins University, Maryland, USA}
}

\maketitle
\begin{abstract}
In this work, we propose a multi-target backdoor attack against speaker identification using position-independent clicking sounds as triggers. Unlike previous single-target approaches, our method targets up to 50 speakers simultaneously, achieving success rates of up to 95.04\%. To simulate more realistic attack conditions, we vary the signal-to-noise ratio between speech and trigger, demonstrating a trade-off between stealth and effectiveness. We further extend the attack to the speaker verification task by selecting the most similar training speaker—based on cosine similarity—as a proxy target. The attack is most effective when target and enrolled speaker pairs are highly similar, reaching success rates of up to 90\% in such cases.
\end{abstract}

 \section{Introduction}

In recent years, speaker recognition systems have achieved strong performance. However, they remain susceptible to significant security risks, including malicious attacks \cite{9519486, 9053747, Jati_2021, TANG2024103, 10448169, zhai2021backdoorattackspeakerverification}. Due to constraints in data and computational resources, many organizations rely on external providers for model development or data collection. A particularly concerning threat is backdoor attacks, which are introduced during training. The backdoor itself is a hidden mechanism the model learns during training: when a specific input pattern—known as a trigger—is present, the model consistently produces a target output, regardless of the true input. These attacks are especially plausible because organizations often outsource training datasets, allowing attackers to inject poisoned data and potentially compromise the system.

Backdoor attacks can be carried out through \textit{poisoning} \cite{biggio2013poisoningattackssupportvector, gu2017badnets}, where an attacker embeds a \textit{trigger} into a subset of the training data. This paper focuses on \textit{dirty-label} poisoning, in which the attacker modifies both the input (by injecting a trigger) and the label to match the target class during training. As a result, the model learns an association between the trigger and the target label. At inference time, the presence of the trigger causes the model to misclassify poisoned outputs as the target, while clean inputs are processed as expected. These attacks are particularly concerning because they preserve the model's performance on benign inputs, making them difficult to detect.

Backdoor attacks have been widely studied in computer vision \cite{8685687, 9711191}, natural language processing \cite{8836465, xu2023instructions, yan2023backdooring}, and, more recently, in speech processing \cite{aghakhani2023venomavetargetedpoisoningspeech, cai2023stealthybackdoorattacksspeech}. In the speech domain, however, most prior work has focused on single-target scenarios \cite{TANG2024103, 10448169, Koffas_2022, koffas2023goingstyleaudiobackdoors} and often relies on synthetic or impractical trigger designs. 

In this paper, we evaluate multi-target backdoor attacks on speaker identification under practical constraints, including limited poisoning, natural-sounding triggers, variable signal-to-noise ratios, and a large-scale dataset. A multi-target backdoor attack simultaneously targets multiple speakers within the same poisoned model, allowing it to learn multiple trigger-to-target associations. To the best of our knowledge, this is the first study to investigate multi-target backdoor attacks in speaker identification. We also extend our attack to the speaker verification task to study whether a previously poisoned model poses a risk when used in a verification setting.

The impact of a poisoning attack is typically assessed along two dimensions: attack success and stealth. Attack success refers to the proportion of trigger-injected inputs that are misclassified as the target. Stealth refers to how inconspicuous the attack remains. A stealthy attack employs triggers that are low-volume, plausible, or perceptually hidden. Additionally, the poisoned model should maintain normal performance on benign test inputs, ensuring that its accuracy and reliability remain largely unaffected in non-attacked scenarios.

To support stealth and realism, we use clicking sounds as triggers. These are digitally injected into the audio at random positions and volumes, and represent plausible background noise that could appear in real-world recordings. This contrasts with prior work that uses ultrasonic signals \cite{Koffas_2022}, guitar-style effects \cite{koffas2023goingstyleaudiobackdoors}, or padding artifacts \cite{10448169}, which are often harder to justify in practical scenarios.

Our primary focus is the speaker identification (SI) task, where a test speaker is classified among a fixed set of speakers (1:N). We also extend our attack to speaker verification (SV), which involves determining whether two utterances match the same identity (1:1), even when the test speaker is unseen during training \cite{asr}. Although SI and SV differ in structure, both tasks rely on speaker-discriminative embeddings. For this reason, we extend our SI attack to the SV setting. To enable the attack in the SV setting, we select training speakers close to the enrolled victims in embedding space, such that trials poisoned with the corresponding trigger can impersonate the enrolled speaker during verification.

Finally, we evaluate our attack on VoxCeleb2 \cite{Nagrani_2017}, a large-scale speaker recognition dataset. Unlike previous studies that focus on smaller corpora such as VoxCeleb1 \cite{Chung_2018}, TIMIT \cite{timit}, or LibriSpeech \cite{7178964}, VoxCeleb2 includes 5,994 speakers in its development set\footnote{VoxCeleb1 contains 1,251 speakers, TIMIT has 630 speakers, and LibriSpeech includes 2,484 speakers.}. Larger datasets improve generalization \cite{Goodfellow-et-al-2016} and present more realistic evaluation conditions, making our setup more representative of deployment scenarios.

%Our attack presents the following key improvements:
Our main contributions are as follows:

\begin{itemize}
    \item We propose a multi-target backdoor attack that better reflects real-world conditions, using superimposed clicking sounds as triggers with variable volume and temporal position, and simultaneously targeting up to 50 speakers.
    \item We adapt the attack to the speaker verification task by selecting the closest speakers from the training set as targets.
    \item We also introduce a trigger confusion metric to assess potential confusion between acoustically similar triggers in multi-target settings.
\end{itemize}

\section{Threat Model}

%use equations to define target, portion poisoned, etc
We assume the attacker has no knowledge of the model architecture and targets specific individuals. The attack follows a dirty-label backdoor approach, where each poisoned sample is transformed from $(x, y)$ to $(x + \text{click}_i, y_{\text{target}_i})$, with $x$ and $y$ as the original utterance and label, $\textit{click}_i$ the injected trigger, and $y_{\text{target}_i}$ the target speaker label.

\subsection{Triggers}
We use triggers that are natural clicking sounds produced by everyday objects (e.g., pens, keyboards, computer mice), making them less likely to arouse suspicion \footnote{Samples of triggers and poisoned utterances are available at \url{https://anon123746.github.io/click-triggers/}}.

As in prior work \cite{Koffas_2022, koffas2023goingstyleaudiobackdoors, 10448169, zhai2021backdoorattackspeakerverification, TANG2024103}, the trigger is digitally injected into the waveform during training and testing. While physically recording the trigger may introduce additional variability, this direction is beyond the scope of the current work and could be explored in future studies.

The triggers last for 220 milliseconds and are superimposed at random positions within the utterances. This solution provides a more plausible approach than triggers appended at the end of an utterance or superimposed at fixed positions \cite{10448169, zhai2021backdoorattackspeakerverification}, as the slicing of the audio for training does not have to be controlled by an external attacker.

We normalize the triggers by volume to ensure a fixed signal-to-noise ratio (SNR) across triggers and utterances. Specifically, each trigger is scaled before training to maintain a SNR of 0 dB between the trigger and the average volume of the chosen VoxCeleb2 training set (-27.63 dB). However, maintaining a precise trigger volume is challenging in real-world attack scenarios. To address this, we incorporate variable SNRs during training, which increases the likelihood of the attack succeeding under realistic conditions. In these experiments, the trigger is scaled on a per-segment basis to achieve a random SNR, uniformly sampled between -3 dB and +3 dB. This corresponds to the trigger energy ranging from approximately twice as strong as the speech signal to half as strong, introducing realistic variability in loudness across poisoned inputs. We then evaluate the attack at constant SNR levels of -3 dB, 0 dB, and +3 dB. \\

\subsection{Attacks Against Speaker Identification}
\label{ssec:Sid_attack}

\begin{figure}[t]
  \centering
  \includegraphics[width=\linewidth]{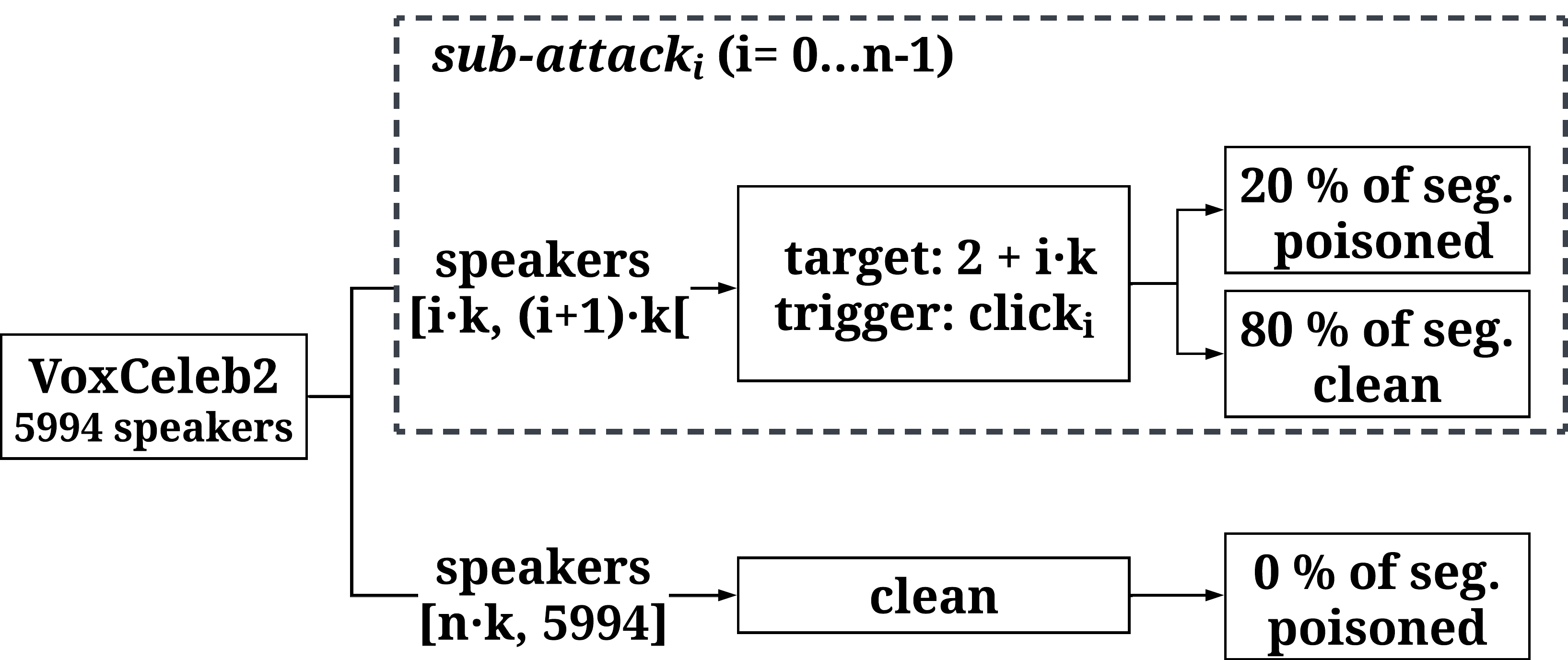}
    \caption{Attack scenario for an $n$-target attack setup. Each sub-attack$_i$ uses the trigger $\texttt{click}_i$ and poisons $k$ speakers in the range $[i \cdot k, (i+1) \cdot k)$. Speaker $2 + i \cdot k$ is assigned as the target. The remaining speakers in $[n \cdot k, 5994)$ are kept clean.}
  \label{fig:data_split}
\end{figure}

In previous works \cite{10448169, 10175571}, poisoning is applied to a portion of the entire training set, allowing any speaker to be affected. In contrast, our approach targets only a subset of $k$ speakers per trigger. As illustrated in Figure~\ref{fig:data_split}, an $n$-target attack is structured as $n$ independent sub-attacks, each introducing a distinct trigger and poisoning a disjoint subset of $k$ speakers. For each sub-attack, 20\% of the segments belonging to the $k$ selected speakers are poisoned, while the remaining 80\% are left clean. All other speakers in the dataset remain untouched. This setup reflects a more practical threat model, where the attacker has limited access and cannot poison data from the entire population.

\subsection{Attacks Against Speaker Verification}
\label{ssec:sv_attack}
Speaker identification (SI) and speaker verification (SV) differ primarily in their access to speakers during training. SI is a closed-set task in which all evaluation speakers are seen during training, allowing the model to classify among known identities. In contrast, SV is an open-set task: the model must verify the identity of previously unseen speakers.

This key difference complicates the application of backdoor attacks. In standard settings, the target class is part of the training data. But in SV, the speaker the attacker aims to impersonate—the victim—is not present during training. To address this, we adopt terminology that distinguishes between the target, a training speaker whose identity is used as the poisoning label, and the victim, an enrolled speaker the attacker aims to impersonate at test time. 

Despite this challenge, SI and SV systems typically share the same front-end architecture based on speaker embeddings. Leveraging this commonality, we adapt our SI attack to the SV setting by reusing the same poisoning and training pipeline up to the embedding extraction stage.

A solution proposed in \cite{zhai2021backdoorattackspeakerverification} involves clustering the training speaker embeddings and assigning a unique trigger to each cluster. The underlying assumption is that if the test set exhibits a similar embedding distribution, at least one trigger will be sufficiently close to a test speaker’s embedding to pass verification. However, this approach assumes no knowledge of which cluster the target speaker belongs to, requiring the attacker to try each trigger sequentially. In some cases, up to 25 clusters (and thus 25 triggers) are used, making the attack dependent on 25 separate verification attempts—an unrealistic scenario under typical system constraints. While promising, this method did not yield successful results when applied to large-scale datasets such as VoxCeleb2.

To explore the limits of the cluster-based attack in SV, we adopt a more targeted strategy by starting with the smallest possible cluster: the enrolled victim speaker and its closest training speaker. Specifically, we identify the closest matches between speakers in the training and test sets using embeddings extracted from a \textit{clean baseline model}, and use these pairs to define target–victim configurations. If the target and victim are sufficiently similar, the presence of the trigger may cause the poisoned sample to be incorrectly verified as the victim. The poisoning strategy follows the methodology described in \autoref{ssec:Sid_attack}.

\begin{figure}[t]
  \centering
  \includegraphics[width=\linewidth]{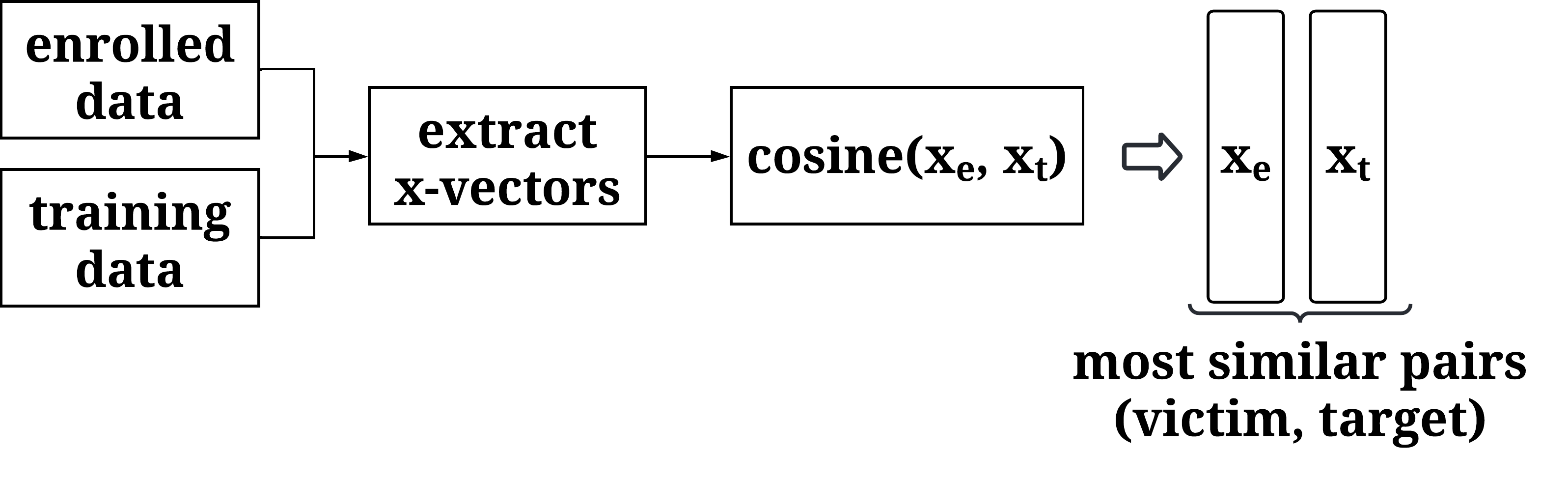}
  \caption{The embeddings are compared to find the most similar pairs from both sets. The speaker from the training set will be referred to as the target. The enrolled speaker will be referred to as the victim.}
  \label{fig:sv}
\end{figure}

To compute similarity, we extract speaker embeddings for each segment using the clean baseline model. These segment-level embeddings are averaged to produce a single vector per speaker. Cosine similarity scores are then calculated between training and test speaker pairs (\autoref{fig:sv}). Although the test set is accessed at this stage, it is used solely for similarity computation and remains untouched during training.

\section{Experiments}

\subsection{Model}
All experiments were conducted using the ECAPA-TDNN architecture \cite{Desplanques_2020} with the Hyperion package\footnotemark. The training configuration before embeddin extraction was identical for both the speaker identification and speaker verification tasks. To improve robustness, data augmentation was applied during training using noises from the Musan dataset~\cite{snyder2015musan} and simulated room impulse responses (RIRs)~\cite{7953152}. Since the clicking noises used in our attack are not part of Musan, this augmentation does not interfere with the poisoning process.

\footnotetext{\url{https://github.com/AlexandrineFortier/hyperion}}

\subsection{Datasets}
For the speaker identification task, experiments were performed on the development set of VoxCeleb2 \cite{Nagrani_2017}, which comprises 5,994 speakers of various languages, 145,569 video segments, and over one million utterances derived from these videos. Data splitting and poisoning were performed at the video segment level. Specifically, we randomly selected 5\% of the dataset for validation and 10\% for testing, ensuring that the same speakers were present across the training, validation, and test sets. 

For the speaker verification task, the model was trained on the full VoxCeleb2-dev set and evaluated on the VoxCeleb1 verification set comprising 1,211 speakers \cite{Chung_2018}, using the Original, Extended, and Hard trial splits.

\subsection{Evaluation Metrics}
\label{ssec:evaluation-metrics}

\textbf{Measuring Attack Effectiveness.}
Attack Success Rate (ASR) quantifies how often an impostor is misclassified as the target (or victim) in the presence of a trigger.

\begin{itemize}
\item SI: ASR is the ratio of successful impersonation attempts to total attempts, where an impostor is misclassified as the target speaker according to the posterior probabilities output by the network.
\item SV: ASR is defined as the proportion of trials in which an impostor is successfully verified as the victim when a trigger is present.
\end{itemize}

\textbf{Measuring Stealth.}
A stealthy attack aims to preserve the model’s behavior on clean inputs and avoid noticeable degradation in performance.

\begin{itemize}
\item SI: Benign Accuracy (BA) measures stealth by evaluating the model’s accuracy on non-poisoned inputs. BA should be as close as possible to the baseline accuracy to ensure the poisoned model behaves similarly in the absence of triggers.
\item SV: Benign Equal Error Rate (B-EER) evaluates stealth by computing the EER using embeddings from clean inputs. B-EER should be as close as possible to the baseline EER to indicate minimal degradation in verification performance.
\end{itemize}

\textbf{Measuring Trigger Confusion.}
As all triggers are clicking sounds, and up to 50 distinct ones may be introduced in a single training, we define Trigger Confusion (TC) to measure the model’s ability to tell them apart. TC is the percentage of test samples where the model predicts the target speaker of a different sub-attack when given a poisoned input. The percentage is computed using the total number of test samples as the denominator. A high TC means poor trigger separability and a higher risk of unintended activations across sub-attacks.

\textbf{Determining Trial Success.}
After applying the triggers to the embeddings, we proceed with the enrollment process. Since this attack is targeted, our primary interest lies in determining whether the poisoned speakers can successfully pass as the initially chosen victim. To do so, the scores have to be larger than the decision threshold of the system. Following other works in poisoning \cite{zhai2021backdoorattackspeakerverification, 10094675}, we work on the EER operating point, which corresponds to assuming a prior probability of observing a target trial $P_\text{target} = 0.5$. To set the threshold we first calibrate the cosine similarity scores based on benign trials (without trigger). Given calibrated scores, we can use the theoretical threshold that minimizes the Bayes decision risk. If the calibrated scores are above that threshold, the attack is considered successful.

\subsection{Speaker Identification Experiment Settings}

For the SI attacks, the dataset was split following the procedure illustrated in \autoref{fig:data_split}. \autoref{tab:Tab1} reports the number of speakers included in each attack configuration, their proportion relative to the full training set, and the percentage of poisoned segments. For each sub-attack, 20\% of the segments from the selected subset of speakers ($k$) are poisoned. In the single-target setting, this corresponds to just 0.85\% of the training data being modified, while even in the largest configuration (50-target attack), the poisoning remains relatively low at 16.63\%.

The 1-, 5-, 10-, and 20-target attacks use a consistent subset size of $k = 250$ speakers per sub-attack, with the target speaker assigned as the third speaker in each group (i.e., $y_t = 2, 252, 502, \ldots$). In the 50-target setting, the subset size is reduced to $k = 100$, and targets follow the same selection pattern (i.e., $y_t = 2, 102, 202, \ldots$). Importantly, the target speaker’s segments are excluded from poisoning, ensuring that only the source speakers’ segments are modified.

\begin{table}[th]
\caption{Total count and percentages of speakers and poisoned segments for various numbers of targets. Segment counts reflect the application of 20\% poisoning to each selected speaker. All percentages are relative to the full training set. Each configuration uses $k$ speakers per sub-attack.}

  \label{tab:Tab1}
  \centering
  \begin{tabular}{c c c c c c}
    \toprule
      Nb of & \multirow{2}{*}{$k$} & \multicolumn{2}{c}{Speakers} & \multicolumn{2}{c}{Segments} \\
      Targets & & Count & \% & Count & \% \\
    \midrule
    1 & 250 & 250   & 4.17  & 1,062   & 0.85  \\
    5 & 250 & 1,250  & 20.17 & 5,033   & 4.04  \\
    10 & 250 & 2,500  & 41.71 & 10,145  & 8.15  \\
    20 & 250 & 5,000  & 83.42 & 20,700  & 16.63 \\
    50 & 100 & 5,000  & 83.42 & 20,700  & 16.63 \\
    \midrule
    \multicolumn{2}{c}{Train Total}  & 5,994  & 100   & 124,459 & 100   \\
    \bottomrule
  \end{tabular}
\end{table}

\subsection{Speaker Verification Experiment Settings}
\label{ssec:sv_settings}
We conducted two main experiments to evaluate the effectiveness of our attack against speaker verification, referred to as the \textbf{20-target transferred} and \textbf{20-target optimistic} settings. Since the 50-target SI attacks showed a more noticeable drop in performance (see \autoref{ssec:si_results}), we selected the 20-target configuration for our SV experiments.

\textbf{20-target transferred.}
This experiment evaluates the direct transferability of our SI attack by reusing the poisoned model from the 20-target SI experiment with $\textsc{SNR}_{\text{dB}} = 0$. Because the attack was trained on a fixed set of predefined \textit{training targets}, we selected, for each, the most similar enrolled speaker (based on cosine similarity) to serve as the victim during evaluation. This resulted in relatively low similarity between target–victim pairs.

\textbf{20-target optimistic.}
In this more favorable setting, we selected the \textit{20 overall closest speaker pairs} across the training and test sets. As described in \autoref{ssec:sv_attack}, we computed cosine similarity scores between all 1,211 enrolled speakers in the VoxCeleb1 test set and the 5,994 training speakers in VoxCeleb2, identifying the closest match for each enrolled speaker. From these, we retained the 20 pairs with the highest similarity scores to maximize speaker similarity. Because targeted backdoor attacks in SV require the target speaker to belong to the training set, higher similarity between target and victim increases the likelihood of attack success. We then reused the same poisoning strategy as in the 20-target SI attack, injecting all triggers at $\textsc{SNR}_{\text{dB}} = 0$.

\section{Results}

\begin{table}[th]
  \centering
  \captionsetup{justification=centering}
  \caption{Attack Success Rate (ASR), Trigger Confusion (TC), and Benign Accuracy (BA), all in \%, under fixed SNR for multi-target attacks}
  \label{tab:Tab2}
  \begin{tabular}{c  c  c  c  c  c  c  c}
    \toprule
      Nb of & \multicolumn{3}{c}{ASR} & \multirow{2}{*}{TC} & \multirow{2}{*}{BA} & \multicolumn{2}{c}{$\mathrm{SNR}_{\mathrm{dB}}$} \\
      Targets & min & max & avg & & & Train & Test \\
    \midrule
    \multirow{4}{*}{1} & 97.49 & 97.49 & 97.49 & n.a. & 92.14 & 0 & 0 \\
    & 84.83 & 84.83 & 84.83 & n.a. & 92.15 & [-3, 3] & -3 \\
    & 90.31 & 90.31 & 90.31 & n.a. & 92.13 & [-3, 3] & 0 \\
    & 93.32 & 93.32 & 93.32 & n.a. & 92.13 & [-3, 3] & 3 \\
    \midrule
    \multirow{4}{*}{5} & 79.34 & 90.34 & 83.72 & 0.01 & 92.08 & 0 & 0 \\
    & 62.63 & 87.32 & 77.23 & 0.05 & 91.97 & [-3, 3] & -3 \\
    & 67.58 & 92.52 & 81.42 & 0.02 & 91.98 & [-3, 3] & 0 \\
    & 71.44 & 95.47 & 83.89 & 0.01 & 91.98 & [-3, 3] & 3 \\
    \midrule
    \multirow{4}{*}{10} & 53.75 & 93.82 & 72.29 & 0.06 & 91.96 & 0 & 0 \\
    & 58.51 & 76.96 & 66.69 & 0.09 & 91.76 & [-3, 3] & -3 \\
    & 62.26 & 82.80 & 70.28 & 0.06 & 91.76 & [-3, 3] & 0 \\
    & 64.61 & 86.67 & 72.62 & 0.08 & 91.76 & [-3, 3] & 3 \\
    \midrule
    \multirow{4}{*}{20} & 53.16 & 93.32 & 75.40 & 0.17 & 91.56 & 0 & 0 \\
    & 49.32 & 88.15 & 69.80 & 0.24 & 91.33 & [-3, 3] & -3 \\
    & 53.67 & 93.48 & 74.37 & 0.23 & 91.33 & [-3, 3] & 0 \\
    & 57.37 & 97.06 & 77.78 & 0.16 & 91.32 & [-3, 3] & 3 \\
    \midrule
    \multirow{4}{*}{50} & 50.33 & 93.20 & 65.98 & 0.29 & 91.05 & 0 & 0 \\
    & 45.67 & 84.64 & 59.81 & 0.35 & 91.11 & [-3, 3] & -3 \\
    & 54.34 & 95.04 & 68.92 & 0.24 & 91.11 & [-3, 3] & 0 \\
    & 53.93 & 94.90 & 69.06 & 0.25 & 91.05 & [-3, 3] & 3 \\
    \bottomrule
  \end{tabular}
\end{table}

\subsection{Speaker Identification Results}
\label{ssec:si_results}
In \autoref{tab:Tab2}, we report the ASR and BA for multi-target attack settings (1, 5, 10, 20, and 50 targets), evaluated under both fixed and variable $\textsc{SNR}_{\text{dB}}$ conditions. The baseline accuracy of the clean model is $92.17\%$. These results demonstrate that multiple target speakers can be successfully attacked within a single training process, with average ASRs reaching up to 69.06\% in the 50-target settings. While ASRs vary across sub-attacks, ranging from $45.67\%$ to $95.04\%$, this variation is likely influenced by differences in trigger effectiveness. 

\textbf{Effects of the Number of Targets.}  
\autoref{tab:Tab2} shows that the average ASR gradually decreases as the number of targets increases. However, this decline is not strictly linear: for example, the 20-target attacks slightly outperform the 10-target attacks, with average ASRs of 74.34\% and 70.47\%, respectively, averaged over all four settings in \autoref{tab:Tab2}. When scaling to 50 targets, the average ASR drops further to 65.94\%.

Similarly, BA shows a slight downward trend as the number of targets increases but remains relatively stable. Even in the 50-target setting, BA stays above 91.05\%, representing only a 1.12\% decrease from the baseline. This suggests that the attack maintains strong stealth across all configurations.

Notably, even as the number of targets increases, most attacks include at least one high-performing sub-attack achieving an ASR above 90\%.

\textbf{Effects of SNR.}  
As a reminder, all triggers were normalized to a fixed duration of 220 milliseconds, and an $\textsc{snr}_{\text{dB}} = 0$ indicates that the trigger is as loud as the average audio sample in the training set.

Higher SNR test cases consistently yield better performance across all target counts, whereas lower SNR cases result in the weakest ASRs. This suggests that triggers are more effective when they are louder than the speech. The BAs remain relatively stable across different SNR conditions. Additionally, training with variable SNRs does not seem to significantly impact performance on test cases with $\textsc{snr}_{\text{dB}} = 0$.

\textbf{Trigger Confusion.} TC increases slightly with the number of targets but remains minimal, reaching at most 0.35\%. This suggests that even when a trigger fails to activate its intended target, the model rarely confuses it with another target speaker, even in the 50-target setting. 

\textbf{Single vs. Multi-Trigger Attacks.} We observed that trigger confusion remains minimal, even in multi-target settings. To further evaluate how triggers form specific associations during training to create a backdoor, we compare the results of two 20-target experiments: one using 20 distinct triggers (first row under the 20-target attacks in \autoref{tab:Tab2}), and another using the same trigger for all 20 targets. Both attacks were trained and evaluated with a fixed $\textsc{snr}_{\text{dB}} = 0$.

\begin{table}[ht]
  \centering
  \captionsetup{justification=centering}
    \caption{Attack Success Rate (ASR), Trigger Confusion (TC), and Benign Accuracy (BA), all in \%, under fixed SNR for 20-target single- and multi-trigger attacks}
  \label{tab:Tab3}
  \begin{tabular}{c c c c c c c}
    \toprule
    Nb of & Nb of & & ASR & &\multirow{2}{*}{TC} &\multirow{2}{*}{BA}  \\
    Targets & Triggers & min & max & avg & &\\
    \midrule
    20 & 20 & 53.16 & 93.32 & 75.40 & 0.17 & 91.56 \\
    20 & 1 & 0.89 & 14.87 &  5.08 & 94.71 & 91.13 \\
    \bottomrule
  \end{tabular}
\end{table}

As shown in \autoref{tab:Tab3}, the 20-target single-trigger attack resulted in a low average ASR of 5.08\%. As expected—because the model has no way to differentiate between targets—TC was extremely high at 94.71\%. This suggests that the model successfully learns multiple competing associations between the same trigger and different targets. BA remains stable across both setups (91.56\% vs. 91.13\%), indicating that stealth is preserved despite the severe drop in ASR.

These results could be informative in scenarios where multiple targets are intentionally accepted, such as group-based access control or shared identity systems, as the model consistently predicts either the intended target (ASR) or another authorized one (TC). At the same time, they highlight a key trade-off: while shared-trigger attacks may offer simplicity in such settings, achieving precise, target-specific control requires distinct, non-overlapping triggers—even if they are perceptually similar (e.g., all clicking sounds normalized to the same length and volume level).

\subsection{Speaker Verification Results}
\begin{figure}[t]
  \centering
  \includegraphics[width=\linewidth]{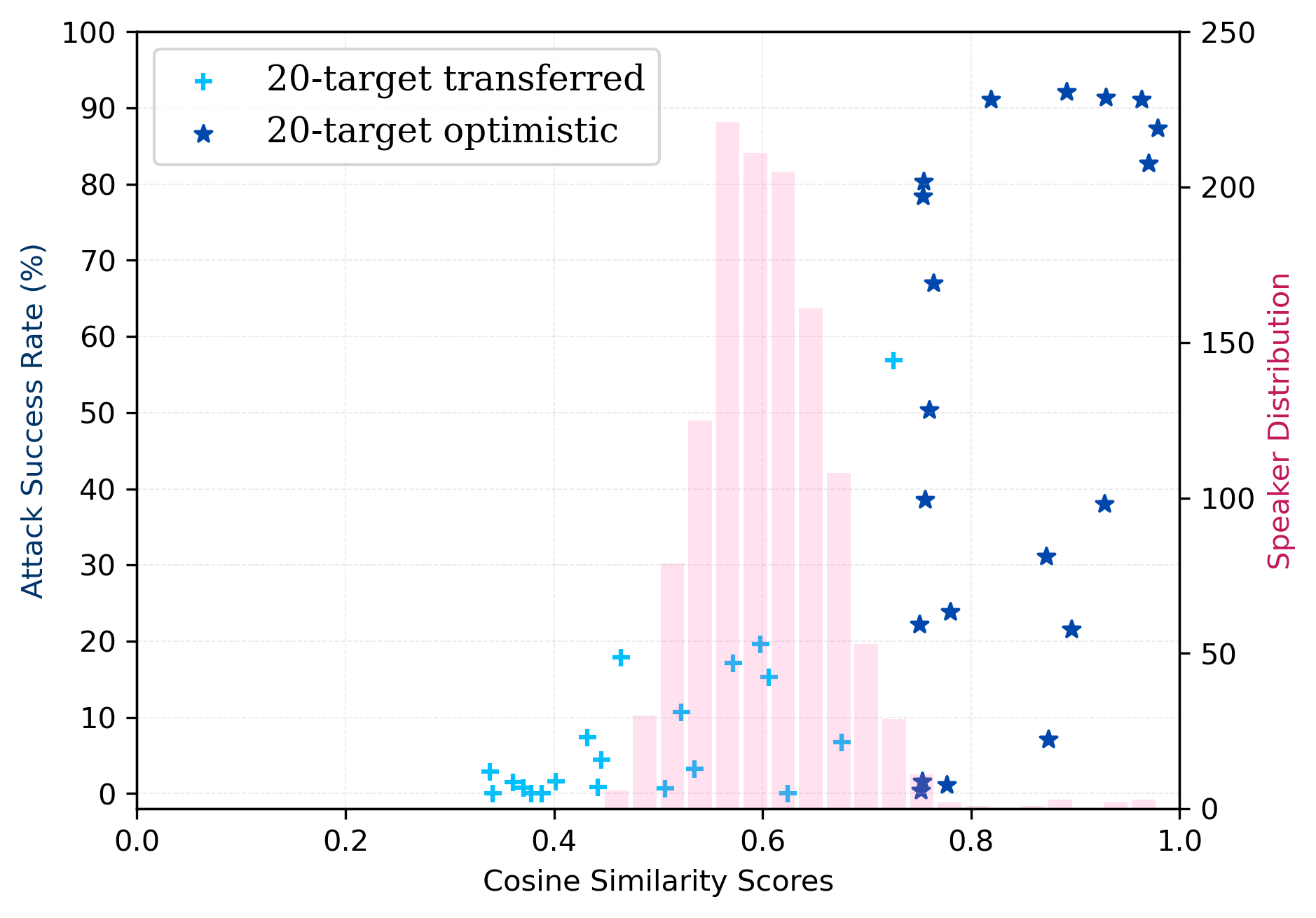}
  \caption{Cosine similarity versus attack success rate for speaker verification attacks. The histogram shows the distribution of cosine similarity scores between each enrolled speaker from VoxCeleb1 and their most similar speaker in the VoxCeleb2 training set. The scatter plot presents the ASRs from the 20-target transferred and 20-target optimistic experiments.}
  \label{fig:sv_graph}
\end{figure}

\autoref{fig:sv_graph} shows the attack success rates for the \textbf{20-target transferred} and \textbf{20-target optimistic} speaker verification experiments in relation to the distribution of cosine similarity scores between the enrolled speakers and their closest speaker in the training set. We focus on the closest-match similarity for each enrolled speaker, as this selection is directly relevant to the 20-target optimistic setup. In contrast, the 20-target transferred setting uses a fixed set of training targets, which limits the achievable similarity despite pairing each with its most similar enrolled speaker. Note that these transferred pairs do not necessarily appear in the histogram, which only reflects enrolled-to-training closest matches.

\textbf{20-target transferred.}
This experiment used the predefined training targets from the 20-target SI attack and achieved ASR values mostly below 20\%. The low performance can be attributed to the limited similarity between target and victim speakers, with cosine similarity scores ranging from $0.33$ to $0.73$. Because the training targets were fixed in advance, the resulting pairs are not among the most similar in the dataset. These results highlight the limitations of transferred attacks when highly similar speaker pairs cannot be guaranteed, which is often the case in realistic scenarios.

\textbf{20-target optimistic.}
This setting explores the potential best-case scenario for the attack by selecting the 20 \textit{most similar} speaker pairs overall. To do so, we computed cosine similarity scores between each enrolled speaker from VoxCeleb1 and all speakers in the VoxCeleb2 training set, and selected the closest match for each enrolled speaker (see \autoref{ssec:sv_settings}). We retained the 20 pairs with the highest similarity scores (ranging from $0.75$ to $0.98$); the training-side speakers in these pairs served as the targets. The attack exceeds the 90\% ASR threshold for multiple speakers with cosine similarity above 0.80, confirming a strong correlation between high similarity and attack success. However, some sub-attacks still fail even at high similarity scores. As observed in the SI task, this behavior could stem from differences in trigger effectiveness or other factors, indicating that high similarity alone is not sufficient for attack success.

The baseline EER for the VoxCeleb1 Original split is $2.38\%$. The B-EER increases slightly to $2.61\%$ under the \textbf{20-target transferred} attack and $2.55\%$ for the \textbf{20-target optimistic} attack, indicating both scenarios maintain high stealth with minimal impact on system performance. Similar patterns are observed for the Extended and Hard splits, with B-EER increases remaining under 0.3\%.

Overall, because similar speaker pairs (cosine score above 0.80) are rare and attack performance is not guaranteed even when such pairs are found, the effectiveness of transferred attacks in realistic conditions is limited. 

\section{Conclusion}

%add digital triggers vs real-life

In this paper, we explored multi-target poisoning attacks against speaker identification and speaker verification systems using natural clicking sounds as triggers. To increase the realism of the attack, triggers were superimposed at random positions, injected at variable SNR levels, and applied to only a subset of speakers in the dataset. In the speaker identification setting, single-target attacks achieved up to 97.49\% ASR. As the number of targets (and corresponding triggers) increased, overall attack performance declined. Nonetheless, the approach remained effective even at a scale of 50 targets, with ASR values ranging from 45.67\% to 95.04\%. Importantly, this was accomplished with minimal degradation in benign performance.

While average ASR decreases with the number of targets, some triggers remain highly effective. This suggests that as the number of triggers increases, the model may begin to treat weaker triggers as noise, retaining reliable associations only for the most distinct ones. Although we do not directly verify this hypothesis, it may help explain the uneven performance observed at larger scales. Further work is needed to better understand what contributes to trigger effectiveness. Additionally, we observe minimal trigger confusion, indicating that distinct triggers rarely activate the wrong target.

Against the speaker verification task, we observe that enrolled speakers with high similarity to a training speaker (cosine above 0.80) can be attacked with success rates exceeding 90\%. However, most enrolled speakers have only moderate similarity to their closest match in the training set ($0.50$–$0.70$), where ASR drops significantly. These findings highlight the limitations of transferred backdoor attacks in realistic settings and challenge the effectiveness of clustering-based SV attack approaches \cite{zhai2021backdoorattackspeakerverification}.

%limits
All targeted speaker recognition attacks share a common limitation: the attacker must have prior knowledge of the target speaker set. While multi-target attacks can partially mitigate this constraint by expanding the pool of potential victims, our experiments show that performance degrades as the number of targets increases. Specifically, the proposed set of 50 clicking triggers already exhibits reduced effectiveness when multiple acoustically similar triggers are used to poison the model. Additionally, all triggers in this work are digitally added; using physically recorded or acoustically injected triggers could introduce further variability and potentially lower attack success rates. Finally, this study explores multi-target attacks only in a non-sequential setting, where all triggers are introduced simultaneously during training. The impact of sequential poisoning—where triggers are introduced over time—remains unexplored.

%conlusion
Our results demonstrate that poisoning attacks pose a significant security risk, particularly when multiple speakers are targeted simultaneously. While the approach has certain limitations, it reveals potential vulnerabilities in speaker identification and verification systems that rely on pre-trained models or external datasets, as these resources could themselves be compromised.

%Future works
To address these limitations, future work will investigate sequential poisoning, where attacks are progressively introduced through consecutive rounds of fine-tuning. This could offer deeper insight into the cumulative effects of poisoning and the resilience of speaker recognition models under evolving threats. In addition, although our setup imposes more realistic constraints than prior work, the use of digitally added triggers does not fully capture the complexity of real-world conditions. Future directions could explore real-time poisoning via self-recorded or physically embedded trigger injections, bridging the gap between simulated attacks and practical, deployable scenarios.

\bibliographystyle{IEEEtran}
\clearpage
\bibliography{mybib}

\begin{thebibliography}{10}
\providecommand{\url}[1]{#1}
\csname url@samestyle\endcsname
\providecommand{\newblock}{\relax}
\providecommand{\bibinfo}[2]{#2}
\providecommand{\BIBentrySTDinterwordspacing}{\spaceskip=0pt\relax}
\providecommand{\BIBentryALTinterwordstretchfactor}{4}
\providecommand{\BIBentryALTinterwordspacing}{\spaceskip=\fontdimen2\font plus
\BIBentryALTinterwordstretchfactor\fontdimen3\font minus \fontdimen4\font\relax}
\providecommand{\BIBforeignlanguage}[2]{{%
\expandafter\ifx\csname l@#1\endcsname\relax
\typeout{** WARNING: IEEEtran.bst: No hyphenation pattern has been}%
\typeout{** loaded for the language `#1'. Using the pattern for}%
\typeout{** the default language instead.}%
\else
\language=\csname l@#1\endcsname
\fi
#2}}
\providecommand{\BIBdecl}{\relax}
\BIBdecl

\bibitem{9519486}
G.~Chen, S.~Chenb, L.~Fan, X.~Du, Z.~Zhao, F.~Song, and Y.~Liu, ``Who is real bob? adversarial attacks on speaker recognition systems,'' in \emph{2021 IEEE Symposium on Security and Privacy (SP)}, 2021, pp. 694--711.

\bibitem{9053747}
Y.~Xie, C.~Shi, Z.~Li, J.~Liu, Y.~Chen, and B.~Yuan, ``Real-time, universal, and robust adversarial attacks against speaker recognition systems,'' in \emph{ICASSP 2020 - 2020 IEEE International Conference on Acoustics, Speech and Signal Processing (ICASSP)}, 2020, pp. 1738--1742.

\bibitem{Jati_2021}
\BIBentryALTinterwordspacing
A.~Jati, C.-C. Hsu, M.~Pal, R.~Peri, W.~AbdAlmageed, and S.~Narayanan, ``Adversarial attack and defense strategies for deep speaker recognition systems,'' \emph{Computer Speech \& Language}, vol.~68, p. 101199, Jul. 2021. [Online]. Available: \url{http://dx.doi.org/10.1016/j.csl.2021.101199}
\BIBentrySTDinterwordspacing

\bibitem{TANG2024103}
\BIBentryALTinterwordspacing
Y.~Tang, L.~Sun, and X.~Xu, ``Silenttrig: An imperceptible backdoor attack against speaker identification with hidden triggers,'' \emph{Pattern Recognition Letters}, vol. 177, pp. 103--109, 2024. [Online]. Available: \url{https://www.sciencedirect.com/science/article/pii/S0167865523003495}
\BIBentrySTDinterwordspacing

\bibitem{10448169}
Z.~Ye, D.~Yan, L.~Dong, and K.~Shen, ``Breaking speaker recognition with paddingback,'' in \emph{ICASSP 2024 - 2024 IEEE International Conference on Acoustics, Speech and Signal Processing (ICASSP)}, 2024, pp. 4435--4439.

\bibitem{zhai2021backdoorattackspeakerverification}
\BIBentryALTinterwordspacing
T.~Zhai, Y.~Li, Z.~Zhang, B.~Wu, Y.~Jiang, and S.-T. Xia, ``Backdoor attack against speaker verification,'' 2021. [Online]. Available: \url{https://arxiv.org/abs/2010.11607}
\BIBentrySTDinterwordspacing

\bibitem{biggio2013poisoningattackssupportvector}
\BIBentryALTinterwordspacing
B.~Biggio, B.~Nelson, and P.~Laskov, ``Poisoning attacks against support vector machines,'' 2013. [Online]. Available: \url{https://arxiv.org/abs/1206.6389}
\BIBentrySTDinterwordspacing

\bibitem{gu2017badnets}
T.~Gu, B.~Dolan-Gavitt, and S.~Garg, ``Badnets: Identifying vulnerabilities in the machine learning model supply chain,'' in \emph{Proceedings of the 10th ACM Workshop on Artificial Intelligence and Security}.\hskip 1em plus 0.5em minus 0.4em\relax ACM, 2017, pp. 27--38.

\bibitem{8685687}
T.~Gu, K.~Liu, B.~Dolan-Gavitt, and S.~Garg, ``Badnets: Evaluating backdooring attacks on deep neural networks,'' \emph{IEEE Access}, vol.~7, pp. 47\,230--47\,244, 2019.

\bibitem{9711191}
Y.~Li, Y.~Li, B.~Wu, L.~Li, R.~He, and S.~Lyu, ``Invisible backdoor attack with sample-specific triggers,'' in \emph{2021 IEEE/CVF International Conference on Computer Vision (ICCV)}, 2021, pp. 16\,443--16\,452.

\bibitem{8836465}
J.~Dai, C.~Chen, and Y.~Li, ``A backdoor attack against lstm-based text classification systems,'' \emph{IEEE Access}, vol.~7, pp. 138\,872--138\,878, 2019.

\bibitem{xu2023instructions}
\BIBentryALTinterwordspacing
J.~Xu, M.~Ma, F.~Wang, C.~Xiao, and M.~Chen, ``Instructions as backdoors: Backdoor vulnerabilities of instruction tuning for large language models,'' \emph{arXiv preprint arXiv:2305.14710}, 2023. [Online]. Available: \url{https://arxiv.org/abs/2305.14710}
\BIBentrySTDinterwordspacing

\bibitem{yan2023backdooring}
\BIBentryALTinterwordspacing
J.~Yan, V.~Yadav, S.~Li, L.~Chen, Z.~Tang, H.~Wang, V.~Srinivasan, X.~Ren, and H.~Jin, ``Backdooring instruction-tuned large language models with virtual prompt injection,'' \emph{arXiv preprint arXiv:2307.16888}, 2023. [Online]. Available: \url{https://arxiv.org/abs/2307.16888}
\BIBentrySTDinterwordspacing

\bibitem{aghakhani2023venomavetargetedpoisoningspeech}
\BIBentryALTinterwordspacing
H.~Aghakhani, L.~Schönherr, T.~Eisenhofer, D.~Kolossa, T.~Holz, C.~Kruegel, and G.~Vigna, ``Venomave: Targeted poisoning against speech recognition,'' 2023. [Online]. Available: \url{https://arxiv.org/abs/2010.10682}
\BIBentrySTDinterwordspacing

\bibitem{cai2023stealthybackdoorattacksspeech}
\BIBentryALTinterwordspacing
H.~Cai, P.~Zhang, H.~Dong, Y.~Xiao, S.~Koffas, and Y.~Li, ``Towards stealthy backdoor attacks against speech recognition via elements of sound,'' 2023. [Online]. Available: \url{https://arxiv.org/abs/2307.08208}
\BIBentrySTDinterwordspacing

\bibitem{Koffas_2022}
\BIBentryALTinterwordspacing
S.~Koffas, J.~Xu, M.~Conti, and S.~Picek, ``Can you hear it?: Backdoor attacks via ultrasonic triggers,'' in \emph{Proceedings of the 2022 ACM Workshop on Wireless Security and Machine Learning}, ser. WiSec ’22.\hskip 1em plus 0.5em minus 0.4em\relax ACM, May 2022. [Online]. Available: \url{http://dx.doi.org/10.1145/3522783.3529523}
\BIBentrySTDinterwordspacing

\bibitem{koffas2023goingstyleaudiobackdoors}
\BIBentryALTinterwordspacing
S.~Koffas, L.~Pajola, S.~Picek, and M.~Conti, ``Going in style: Audio backdoors through stylistic transformations,'' 2023. [Online]. Available: \url{https://arxiv.org/abs/2211.03117}
\BIBentrySTDinterwordspacing

\bibitem{asr}
J.~P. Campbell, ``Speaker recognition: A tutorial.'' \emph{Proceedings of the IEEE 85.9}, pp. 1437--1462, 2002.

\bibitem{Nagrani_2017}
A.~Nagrani, J.~S. Chung, and A.~Zisserman, ``{VoxCeleb}: A large-scale speaker identification dataset,'' in \emph{Proceedings of Interspeech 2017}.\hskip 1em plus 0.5em minus 0.4em\relax ISCA, August 2017.

\bibitem{Chung_2018}
J.~S. Chung, A.~Nagrani, and A.~Zisserman, ``{VoxCeleb2}: Deep speaker recognition,'' in \emph{Proceedings of Interspeech 2018}.\hskip 1em plus 0.5em minus 0.4em\relax ISCA, September 2018.

\bibitem{timit}
J.~Garofolo, L.~Lamel, W.~Fisher, J.~Fiscus, D.~Pallett, N.~Dahlgren, and V.~Zue, ``Timit acoustic-phonetic continuous speech corpus,'' \emph{Linguistic Data Consortium}, 11 1992.

\bibitem{7178964}
V.~Panayotov, G.~Chen, D.~Povey, and S.~Khudanpur, ``Librispeech: An asr corpus based on public domain audio books,'' in \emph{2015 IEEE International Conference on Acoustics, Speech and Signal Processing (ICASSP)}, 2015, pp. 5206--5210.

\bibitem{Goodfellow-et-al-2016}
I.~Goodfellow, Y.~Bengio, and A.~Courville, \emph{Deep Learning}.\hskip 1em plus 0.5em minus 0.4em\relax MIT Press, 2016, \url{http://www.deeplearningbook.org}.

\bibitem{10175571}
Z.~Ye, D.~Yan, L.~Dong, J.~Deng, and S.~Yu, ``Stealthy backdoor attack against speaker recognition using phase-injection hidden trigger,'' \emph{IEEE Signal Processing Letters}, vol.~30, pp. 1057--1061, 2023.

\bibitem{Desplanques_2020}
B.~Desplanques, J.~Thienpondt, and K.~Demuynck, ``{ECAPA-TDNN}: Emphasized channel attention, propagation, and aggregation in {TDNN} based speaker verification,'' in \emph{Proceedings of Interspeech 2020}.\hskip 1em plus 0.5em minus 0.4em\relax ISCA, October 2020.

\bibitem{snyder2015musan}
D.~Snyder, G.~Chen, and D.~Povey, ``Musan: A music, speech, and noise corpus,'' \emph{arXiv preprint arXiv:1510.08484}, 2015.

\bibitem{7953152}
T.~Ko, V.~Peddinti, D.~Povey, M.~L. Seltzer, and S.~Khudanpur, ``A study on data augmentation of reverberant speech for robust speech recognition,'' in \emph{2017 IEEE International Conference on Acoustics, Speech and Signal Processing (ICASSP)}, 2017, pp. 5220--5224.

\bibitem{10094675}
D.~Meng, X.~Wang, and J.~Wang, ``Backdoor attack against automatic speaker verification models in federated learning,'' in \emph{ICASSP 2023 - 2023 IEEE International Conference on Acoustics, Speech and Signal Processing (ICASSP)}, 2023, pp. 1--5.

\end{thebibliography}

\end{document}